\documentclass[a4paper,12pt]{article}

\newif\ifPDFLaTeX
\expandafter\ifx\csname pdfoutput\endcsname\relax\else\PDFLaTeXtrue\fi
 \ifPDFLaTeX
  \usepackage[pdftex,colorlinks]{hyperref}
  \usepackage[pdftex]{graphicx}
\else
  \usepackage{graphicx}
\fi
\providecommand{\preprintno}[1]{\relax}
\newcommand{\graphicswidth}{0.75\textwidth}
\usepackage{fancyhdr}
\def\preprint#1{\gdef\thepreprint{#1}}
\def\thepreprint{}
\fancypagestyle{fancyplain}{
  \fancyhf{}
  \fancyhead[R]{\thepreprint \\\hfill\\ June 10, 2005}
  
  \addtolength{\headsep}{3.5cm}
}
\usepackage{amsmath,amssymb}
\usepackage{cite}
\providecommand{\href}[2]{#2}
\makeatletter
\renewcommand\section{\@startsection {section}{1}{\z@}%
                                   {-3.5ex \@plus -1ex \@minus -.2ex}%
                                   {2.3ex \@plus.2ex}%
                                   {\large\bf}}   
\renewcommand\subsection{\@startsection {subsection}{1}{\z@}%
                                   {-3.5ex \@plus -1ex \@minus -.2ex}%
                                   {2.3ex \@plus.2ex}%
                                   {\normalfont\bf}}         
\makeatother
\newcommand{\ii}{\mathrm{i}}

\newcommand{\qed}{\mathrm{QED}}
\makeatletter
\newcommand{\fmslash}[2][0mu]{%
  \mathchoice
    {\fmsl@sh\displaystyle{#1}{#2}}%
    {\fmsl@sh\textstyle{#1}{#2}}%
    {\fmsl@sh\scriptstyle{#1}{#2}}%
    {\fmsl@sh\scriptscriptstyle{#1}{#2}}}
\newcommand{\fmsl@sh}[3]{%
  \m@th\ooalign{$\hfil#1\mkern#2/\hfil$\crcr$#1#3$}}
\makeatother
\allowdisplaybreaks
\numberwithin{equation}{section}

\begin{document}

\preprint{MZ-TH/05-07\\ hep-ph/0504240 }

\title{Unitarity constraints on top quark \\ signatures of Higgsless models}
\author{ \normalsize
Christian Schwinn\thanks{schwinn@thep.physik.uni-mainz.de}\\
{\normalsize\it Institut f\"ur Physik, Johannes-Gutenberg-Universit\"at}\\
{\normalsize\it Staudingerweg 7,  D-55099 Mainz, Germany}}
\setcounter{tocdepth}{2}
\setcounter{secnumdepth}{3}
 \date{}

\maketitle
\thispagestyle{fancyplain}
\setcounter{page}{0}
\begin{abstract}
We use conditions for unitarity cancellations 
to constrain the couplings of the top and
bottom quarks to Kaluza-Klein modes
 in Higgsless models of electroweak symmetry breaking.
An example for the mass spectrum of quark resonances in a theory space
model is given
and the implications for the collider phenomenology in the top sector
are discussed, comparing to signatures of
Little Higgs and strong electroweak symmetry
breaking models.\\[1cm]
\end{abstract}

\newpage

\section{Introduction}
Recently it has been suggested that a viable description of
electroweak symmetry breaking (EWSB) without a Higgs boson, weakly
coupled up to energies of $\Lambda=5$-$10$ TeV, may be achieved
employing gauge symmetry breaking by boundary conditions in a higher
dimensional spacetime~\cite{Csaki:2003dt}.  Variants of the setup in a
warped or flat extra dimension~\cite{Csaki:2003zu} and four
dimensional `theory space'
models~\cite{Foadi:2003xa,Chivukula:2004pk,Georgi:2004iy} have been
constructed. The mechanism of gauge symmetry breaking by boundary
conditions has been investigated from several
perspectives~\cite{OS:SR,Nagasawa:2004xk} and applied to extended
gauge symmetries and six dimensional models~\cite{Gabriel:2004ua}.
The basic idea of the higher dimensional Higgsless models is to cure
the bad high energy behavior of scattering amplitudes of massive gauge
bosons by the exchange of Kaluza-Klein (KK)-excitations of the
electroweak gauge bosons.  The relevant unitarity cancellations are
ensured by higher dimensional gauge
invariance~\cite{Csaki:2003dt,OS:SR,SekharChivukula:2001hz} or by a
large number of sites in theory space models~\cite{Chivukula:2002ej}.
A cutoff scale $\Lambda$---reflecting the non-renormalizable nature of
the higher dimensional gauge theory---is implied by partial wave
unitarity bounds~\cite{SekharChivukula:2001hz,Chivukula:2002ej,Papucci:2004ip}
resulting from the increasing number of open channels.

Fermion masses in 5D Higgsless models can be generated by allowing
the fermions to propagate
in the extra dimension and introducing appropriate boundary localized kinetic
and mass terms~\cite{Csaki:2003sh} which is consistent with unitarity
cancellations as long as the reduced gauge symmetry at the boundaries is
respected by the localized terms~\cite{CS:HLF}.

Early phenomenological
studies of the five dimensional models considered fermions localized near a
brane and found it difficult to accommodate electroweak precision
data~\cite{Burdman:2003ya,Davoudiasl:2003me}
while raising the cutoff scale 
significantly compared to four dimensional strongly interacting models~\cite{Papucci:2004ip}, in agreement with theory space results~\cite{Chivukula:2004pk,Georgi:2004iy}.
However, as pointed out
subsequently~\cite{Cacciapaglia:2004rb,Foadi:2004ps} the delocalization
of the fermions in the extra dimension 
or analogous theory space constructions~\cite{Chivukula:2005bn,Casalbuoni:2005rs,Chivukula:2005xm} allows to  suppress the 
couplings of the light fermions to the KK-gauge bosons and 
might be the key for a realistic model.

While no fully realistic model has emerged yet, recent
studies~\cite{Chanowitz:2004gk,Birkedal:2004au} concentrate on
signatures of the mechanism of Higgsless EWSB in gauge boson
scattering at the CERN Large Hadron Collider (LHC).  To identify
generic signatures in this channel, ref.~\cite{Birkedal:2004au}
utilizes the conditions for unitarity
cancellations~\cite{Csaki:2003dt,OS:SR} to constrain the couplings of
the gauge boson KK-resonances.  The aim of the present work is to
constrain signatures of Higgsless models in the top quark sector using
a similar approach. We show that the KK-excitations of the third
family quarks are also a \emph{generic} feature of a Higgsless model
which remains perturbative up to a scale $\Lambda \approx 5$-$10$ TeV,
although the detailed structure of the mass spectrum and the coupling
constants is more model dependent than in gauge boson scattering.

The large top quark mass implies that the third family quarks play a
special role in Higgsless models, for instance it appears difficult to
delocalize the third family in the same way as the lighter
fermions~\cite{Cacciapaglia:2004rb}.  In addition, it has been
argued~\cite{Cacciapaglia:2004rb} that precision electroweak
constraints on the $Zb\bar b$ vertex require the first KK-excitations
of the third family quarks to be considerably heavier than the gauge
boson KK-modes. In this context it is interesting to recall the
Appelquist-Chanowitz~(AC) unitarity
bound~\cite{Appelquist:1987cf,Golden:1994pj} $\Lambda_t
=8\pi v^2/(\sqrt {2 N_c }m_t) \sim 3.5$ TeV on the the scale of mass
generation for the top quark (we have used the slightly tighter bound
given recently by Dicus and He in~\cite{Golden:1994pj}).
 Hence there is a potential tension between
the AC bound and large masses of top-quark KK-modes.  On the other
hand, this bound suggests that signals of the mechanism of mass
generation are more likely to be observable in the third family quarks
than for lighter fermions, especially if the light fermions are
delocalized and decouple from the gauge boson KK-modes.

In section~\ref{sec:examples} we review previous results on third
family quarks in 5D Higgsless models and give an example for the mass
spectrum of the KK bottom and top quarks in a theory space 
model.  In section~\ref{sec:sr} we will present the sum rules ensuring
the boundedness of the 4 particle scattering amplitudes involving
third family quarks.  Based on this sum rules, in
section~\ref{sec:pheno} we introduce a simple scenario for the first
KK-level of gauge bosons and third family quarks in a Higgsless model
and comment on the resulting collider phenomenology, comparing it to
that of the heavy top quark in Little Higgs models and of heavy gauge
bosons in models of strong EWSB.

\section{Third family quarks in Higgsless models}
\label{sec:examples}
In this section we recall different possibilities for the
 fermionic sector of Higgsless models  and the resulting
KK-spectrum in the top quark sector (For convenience, the
term `KK-mode' will be used also in theory space models). 
After a short overview over 5D models, in subsection~\ref{sec:deconstruct}
a simple theory space  model is discussed in some more detail.
While it is beyond the
scope of the present work to construct a fully realistic model, 
this section provides us with an example for a
 mass spectrum of the KK modes that
will be used for purposes of illustration later on.  
Some aspects of the
top quark sector in Higgsless models with a warped extra dimension
have been discussed in~\cite{Burdman:2003ya,Cacciapaglia:2004rb}.

Let us briefly recall the setup of 5D models and the features of the
fermion mass spectrum.  The models of~\cite{Csaki:2003dt,Csaki:2003zu}
employ a left-right symmetric bulk gauge group $SU(2)_L\otimes
SU(2)_R\otimes U(1)$.  The same group structure is also used in warped
models including a Higgs scalar~\cite{Agashe:2003zs}. Flavor physics
in these models has been discussed in~\cite{Agashe:2004ay}.  On the
brane at $y=\pi R\equiv\ell$ the left-right symmetry is broken to the
diagonal subgroup $SU(2)_{L+R}$, on the second brane at $y=0$ the
symmetry breaking pattern is $SU(2)_R\otimes U(1)\to U(1)_Y$.  Left-
and right-handed fermions arise as zero modes of bulk $SU(2)_L$ and
$SU(2)_R$ doublets $\Psi_L$ and $\Psi_R$, respectively.  

Without
additional structure, the zero mode fermions are massless and there
are degenerate Dirac-KK modes for the left- and right-handed fermions.
To give masses to the zero modes, Dirac masses consistent with the
unbroken $SU(2)_{L+R}$ can be added on the brane at
$y=\ell$~\cite{Csaki:2003sh}.  To obtain a mass splitting between the
up and down type quarks, two different possibilities have been
considered.  In~\cite{Csaki:2003sh} the right-handed up- and down type
quarks are contained in the same bulk $SU(2)_R$ doublet $\Psi_R$. To
lift the mass degeneracy in the isospin multiplets, one can use the
$y=0$ brane where the broken $SU(2)_R$ allows to add boundary kinetic
terms for the right handed down-type quarks. The large boundary term
needed in this setup to obtain the mass splitting between the top and
bottom quark also results in a large mass splitting of the first
KK-modes of bottom and top quark. In the setup of~\cite{Agashe:2003zs}
also used in a Higgsless model in~\cite{Burdman:2003ya}, two $SU(2)_R$
doublets $\Psi^t_R$ and $\Psi^b_R$ are introduced that contain the
right-handed top and bottom quarks as zero modes.  This allows to use
different brane masses for top and bottom quarks at $y=\ell$ so no
large boundary kinetic term is needed to split the isospin doublets.
Additional structure must be added on the $y=0$ brane to give large
masses to the unwanted down-type zero mode of $\Psi^t_R$ and 
the up-type zero mode of $\Psi^b_R$.

The large boundary mass term necessary for the top quark mass
 splits the initially degenerate KK-modes of $\Psi_L$ and $\Psi_R$.
  In a flat extra-dimension, the splitting
turns out to be similar to the brane mass itself, in a warped extra
dimension the effects can be even larger depending on the localization
of the zero mode fermions~\cite{Burdman:2003ya,Chang:2005vj}.  Thus,
in this setup the mass of the lightest top quark KK-mode is likely to
be significantly {\it lighter} than that of the gauge bosons.  Since
perturbativity in gauge boson scattering requires the gauge boson
KK-modes to be as light as possible~\cite{Papucci:2004ip}, the first
KK-mode of the top quark would therefore be dangerously light.
In the setup with $t_R$ and $b_R$ in the same bulk doublet $\Psi_R$~\cite{Csaki:2003sh}, the large boundary kinetic term needed to get the top-bottom mass
splitting will make the first bottom KK-mode even lighter, 
while in the setup with separate $\Psi^t_R$ and $\Psi^b_R$ doublets, 
one expects only a small mass splitting of the bottom KK-modes. 
Furthermore, reference~\cite{Cacciapaglia:2004rb} finds a tension in
obtaining a large enough top mass while obeying constraints on the
$Zbb$ vertex and concludes that in a realistic warped 5D model the
masses in the top quark sector must be generated by a separate
mechanism.  This tension is reduced if one allows the gauge boson
KK-modes to become heavier by giving up the demand for perturbativity
up to $5$-$10$ TeV~\cite{Burdman:2003ya}.  Also warped models
including a a Higgs scalar~\cite{Agashe:2003zs,Agashe:2004ay} and with
the first gauge boson KK level at $3$-$4$ TeV are less affected by
this problem.

\subsection{Theory space setup}
\label{sec:deconstruct}
As an example for a theory space Higgsless model including delocalized fermions,  we  consider the `one-site delocalized' setup 
of ref.~\cite{Chivukula:2005bn} that performed
a numerical analysis for
a 4-site model. We will now extend this simple model 
 to include right-handed fermions and
generate fermion masses\footnote{Recently a similar
  construction for fermions delocalized over an arbitrary number of
  sites has been given~\cite{Chivukula:2005xm} but the zero modes were
  treated as massless.  Ref~\cite{Casalbuoni:2005rs} uses another
  approach with localized fermions coupling nonlocally to the vector
  bosons. While this setup improves agreement with electroweak
  precision data, the unitarity issue is not addressed since no heavy
  partners for the top- and bottom quarks are introduced.}.
We thus consider a $SU(2)^3\otimes U(1)$ gauge theory with $3$ nonlinear sigma
models with symmetry breaking pattern $SU(2)\otimes SU(2)/SU(2)$ acting as
link fields.
The link fields $U_i$ 
transform under the gauge transformations as
\begin{equation}
  U_i\to g_{i-1}^\dagger U_i g_i
\end{equation}
where $i\in \{1,2,3\}$.  Here for $i<3$ the gauge transformations $g_i$ are
elements of $SU(2)_{i}$  while for $i=3$
there is only a $U(1)$ gauge transformation.

In~\cite{Chivukula:2005bn} a left handed fermion doublet
$\psi_L=(\psi_{t,L},\psi_{b,L})$ is introduced on the site $i=0$ and a
doublet of vector-like fermions $\Psi_1$ on the second site $i=1$.
Both have the same Hypercharge and therefore are charged under the
$U(1)$ on site $3$.  The continuum limit will be similar to the setup
of~\cite{Foadi:2004ps} based on a single bulk $SU(2)$ where also
non-local couplings of the fermions to the unbroken $U(1)$ located on
one boundary have to be introduced.  Avoiding this nonlocality in
theory space requires additional $U(1)$ groups, corresponding to a
faithful deconstruction of the 5D $SU(2)_R\otimes SU(2)_L\otimes U(1)$
5D model~\cite{Chivukula:2004pk} which will not be considered in the
following.  To give mass to the standard model~(SM) fermions, we will
additionally include two right-handed fermions $\chi_{bR}$ and
$\chi_{tR}$ on the $U(1)$ site, with the same Hypercharges as the
right-handed SM up- and down-type quarks.  Finally, a second
vector-like doublet $\Psi_2$ is introduced at the site $i=2$.  A gauge
invariant Lagrangian is given by
\begin{multline}\label{eq:ts-masses}
  -\mathcal{L}_M= M_{\Psi 1} \bar\Psi_{1L}\Psi_{1R}+ M_{\Psi 2} \bar\Psi_{2L}\Psi_{2R} 
 + y_1 f_1 \bar \psi_L U_1 \Psi_{1R} +  y_2 f_2 \bar \Psi_{1L} U_2 \Psi_{2R}\\
  + y_3 f_3 \bar \Psi_{2L} U_3 (\lambda^0+\lambda^3\sigma^3) \chi_{R}
+\,\text{h.c.}
\end{multline}
where the notation $\chi_{R}=(\chi_{tR},\chi_{bR})$ has been introduced.
Here `nonlocal' gauge invariant terms involving products of the link
fields have been discarded. They are usually argued to be generated by higher order effects
only and therefore suppressed.
The mass matrix for the top quark sector is given by
\begin{equation}
   -\mathcal{L}_{M_t}= 
   \begin{pmatrix}
     \bar \psi_t,& \bar \Psi_{t1}, &\bar \Psi_{t2}
   \end{pmatrix}_L
   \begin{pmatrix}
     f_1 y_1 &0 &0\\
     M_{\Psi 1}&f_2 y_2&0\\
     0 & M_{\Psi 2}& f_3 y_t
   \end{pmatrix}
\begin{pmatrix}
      \Psi_{t1}\\\Psi_{t2} \\ \chi_t
   \end{pmatrix}_{R}
+\,\text{h.c.}
\end{equation}
with $y_t=y_3 (\lambda^0+\lambda^3)$. The matrix for the bottom sector
differs only in the third  entry on the diagonal that is instead given by
 $y_b=y_3 (\lambda^0-\lambda^3)$.
 
 For the `one-site delocalized' model without $\chi_R$ and $\Psi_2$, there is
 a massless zero mode $\psi_{L}^{(0)}=(\cos\theta_{\Psi 1} \psi_L
 -\sin\theta_{\Psi 1}\Psi_{1L})$ with $\sin \theta_{\Psi 1} =f_1
 y_1/\sqrt{(f_1 y_1)^2+M_{\Psi 1}^2}$.  Applying this setup to the light
 fermions, good agreement with electroweak precision data was found
 in~\cite{Chivukula:2005bn} for the parameters $f_1=300$ GeV and $f_2=f_3=592$
 GeV and a mixing angle $ \sin^2\theta_{\Psi 1}=0.01477$ that translates into
\begin{equation} \label{eq:mpsi1}
  M_{\Psi_1}=\frac{f_1 y_1}{\tan\theta_{\Psi 1}}= y_1\times 2450 \, \text{GeV}
\end{equation}
while the masses of the KK gauge boson are given
by~\cite{Chivukula:2005bn} $m_{W^{(1)}}\sim m_{Z^{(1)}}\sim 890$ GeV.
Remarkably, already in this simple model the `KK excitation'
$\psi_{L}^{(1)}=(\sin\theta_{\Psi 1} \psi_L +\cos\theta_{\Psi
  1}\Psi_{1L})$ has a mass of 
$m_{\psi^{(1)}}=\sqrt{(y_1f_1)^2+ M_{\Psi 1}^2}=y_1 \times 2.5$ TeV
and is naturally heavier than the gauge boson KK-modes.  For purposes
of illustration, we use the same input parameters for our extended
setup for the third family quarks.  As an example for a mass spectrum
where the mass splitting of the first top and bottom quark masses is
not too large, for the parameters $y_1=1.5$, $y_2=4$, $y_t=5$,
$y_b=0.08$ and choosing $M_{\Psi 1}=M_{\Psi 2}$ as in~\eqref{eq:mpsi1}
one obtains the spectrum
\begin{equation}
  \begin{aligned}
     m_t&=178.7 \,\text{GeV} & m_{T^1}&= 3.1\,\text{TeV}& m_{T^2}&=5.6\,\text{TeV} \\
    m_b&=4.6 \, \text{GeV} &m_{B^1}&=2.5  \,\text{TeV}&m_{B^2}&=5.5 \, \text{TeV}
  \end{aligned}
\end{equation}
In contrast to the five dimensional setup, there are no almost
degenerate KK-modes in this model.  From the eigenvectors of the
square of the mass matrix in~\eqref{eq:ts-masses} one finds 
that---because of the large Yukawa coupling $y_t$---the right handed top-quark
zero mode has a considerable admixture of $\Psi_{t 1/2,R}$ while the
other zero modes are approximately `localized' on the sites $0$
and $3$. Such a composite structure of the right-handed top
is also expected in warped 5D models, where the
right-handed top quark has to be localized near the brane where EWSB
takes place~\cite{Burdman:2003ya,Cacciapaglia:2004rb,Agashe:2004ay}.
Thus, corrections to the right-handed top quark couplings seem to be
another generic prediction of the class of Higgsless models considered
here, in addition to the direct signatures of the KK-modes that are
the focus of the remainder of this work.

\section{Unitarity cancellations in the top quark sector}
\label{sec:sr}
To introduce our method and to present results needed in the following,
we briefly review the unitarity bounds on the couplings of the
KK-gauge bosons to the $W$ and $Z$ obtained in~\cite{Birkedal:2004au}.
The cancellation of terms growing like $E^4$ and $E^2$ in
the $WZ\to WZ$ and $W^+ W^-\to W^+ W^-$ scattering amplitudes implies
the sum rules (see also~\cite{Csaki:2003dt,LlewellynSmith:1973,OS:SR})
\begin{equation}
\label{eq:wz-sr}
\begin{aligned}
 3 \sum_n  g^2_{ZWW^{(n)}}m^2_{W^{(n)}}&=   g^2_{ZWW} \frac{ m_Z^4}{m_W^2}\\
   3 \sum_n g_{WWZ^{(n)}}^2 m_{Z^{(n)}}^2&=
 \left(4m_W^2-3 m_Z^2\right)g_{WWZ}^2+ 4m_W^2g_{WW\gamma}^2
\end{aligned}
\end{equation}
up to terms suppressed by an order of $(m_W
/m_{W^(n)})^2$.
Therefore  upper bounds~\cite{Birkedal:2004au} on the interaction of 
the $W$ and $Z$ KK-modes can be obtained:
\begin{equation}\label{eq:bmp}
  \begin{aligned}
g_{ZWW^{(1)}}&\lesssim \frac{g_{ZWW}m_Z^2}{\sqrt 3 m_{W^{(1)}} m_W}  
\overset{\text{SM}}{=}\frac{g m_Z}{\sqrt 3 m_{W^{(1)}}}\\ 
     g_{WWZ^{(1)}}&\overset{\text{SM}}{\lesssim}\frac{g m_W}{\sqrt 3 m_{Z^{(1)}}} 
  \end{aligned}
\end{equation}
We have indicated the result of
inserting the tree level SM values for the zero mode couplings.
 One expects corrections
to these values in Higgsless models~\cite{Csaki:2003dt} but 
the SM values will be used for purposes of illustration.
In the remainder of this section, the relations corresponding
to~\eqref{eq:wz-sr} in the top-quark sector are discussed, the
constraints on the coupling constants corresponding to~\eqref{eq:bmp}
will be subject of section~\ref{sec:pheno}. 
\subsection{Unitarity sum rules for third family quarks}
 Consider the scattering
of zero mode fermions, denoted by $q_i$, and zero mode vector bosons,
denoted by $V_a$, that couple to the corresponding KK-modes
$Q_i^{(n)}$ and $V_a^{(n)}$.  We will only be concerned with
interactions that involve a single KK-excitation and parameterize the
corresponding interaction Lagrangian as
\begin{multline}\label{eq:lagrangian}
  \mathcal{L}_{\text{int}}= \sum_{n} \Bigl\{
  \ii g_{V_aV_bV_c^{(n)}} \partial^\mu V_a^\nu V_{b\mu} V_{c\nu}^{(n)}\\
+\left[
g^{L/R}_{i(j,n)V_a}\bar q_i \fmslash V_a 
\left(\tfrac{1\pm\gamma^5}{2}\right)Q_j^{(n)}
+g^{L/R}_{ijV_a^{(n)}}\bar q_i\fmslash V_a^{(n)}
\left(\tfrac{1\pm\gamma^5}{2}\right)
  q_j \right]+ \,\text{h.c.}\Bigr\}
\end{multline}
where we have included the zero modes in the sums by defining $Q^{(0)}_i=q_i$ 
and similarly for the vector bosons. A sum over the internal quantum  numbers
is implied. With the convention of~\eqref{eq:lagrangian},
 in the SM we have for instance $g_{ZW^+W^-}=g \cos\theta_w$ where $g=2m_W/v$ is the weak coupling constant.

The unitarity sum rules for this situation can be obtained in a
straightforward way from the results of~\cite{LlewellynSmith:1973} by
omitting the Higgs contributions and allowing for an infinite number
of intermediate particles. 
The cancellation of the leading divergences $\propto E^2$
in the scattering $\bar q_i q_k\to V_a V_b$
is ensured by the sum rule.
\begin{subequations}\label{eq:fermi-srs}
\begin{equation}\label{eq:fermi-lie}
\sum_{n,j} \left[ g^{L/R}_{i (j,n) V_a} g^{L/R}_{(j,n)k V_b}
-g^{L/R}_{i (j,n) V_b} g^{L/R}_{(j,n)k V_a}\right ]
=\sum_{n,c} g_{V_aV_b V_c^{(n)}}g^{L/R}_{i k V_c^{(n)}}
\end{equation}
This is just the generalization of the Lie algebra
$[\tau^a,\tau^b]=\ii f^{abc}\tau^c$ that holds in a gauge theory where
the fermions live in a representation of the gauge group generated by
the $\tau^a$ and the structure constants $f^{abc}$ enter the triple
gauge boson vertex.  This relation therefore is not associated with
the symmetry breaking mechanism.  In contrast, for the cancellation of
the subleading divergences $\propto E$ usually a Higgs boson is
invoked. In absence of Higgs scalars, this sum rule takes the form
\begin{multline}\label{eq:fermi-gold-KK}
\sum_{n,j}  m_{q_i} g^L_{i (j,n) V_b } g^L_{ (j,n) k V_a}
+m_{q_k} g^R_{i (j,n)V_a } g^R_{ (j,n) k V_b}
 -m_{q_j^{(n)}} \left(g^R_{i (j,n) V_b } g^L_{ (j,n)k V_a}        +
 g^R_{i (j,n)V_a}  g^L_{(j,n)kV_b} \right)\\
 = \sum_{c,n}
\frac{(m_{V_a}^2-m_{V_b}^2-m_{V_c^{(n)}}^2)}{2m_{V_c^{(n)}}^2}
g_{V_aV_bV_c^{(n)}} \left(m_{q_i} g^L_{i k V_c^{(n)}}-m_{q_k} 
g^R_{i k V_c^{(n)}} \right)
\end{multline}
\end{subequations}
and the same equation with left- and right-handed couplings exchanged. 
It can be shown that in the special case of identical
fermions and bosons as external particles, the right-handed version
of~\eqref{eq:fermi-gold-KK} is not an independent condition 
but is satisfied identically after~\eqref{eq:fermi-lie}
is used, in agreement with a result of Gunion~{\it et.al} 
in~\cite{LlewellynSmith:1973}.

The  relations
\eqref{eq:fermi-srs} are a consequence of the
underlying higher dimensional gauge symmetry and can consequently also
be obtained from the Ward Identities of the theory~\cite{OS:SR}.  As
verified in~\cite{OS:SR,CS:HLF}, these relations are left intact by
gauge symmetry breaking by Dirichlet boundary conditions and boundary
terms for the fermions consistent with the reduced gauge symmetry on
the brane.  Similar to the case of gauge boson
scattering~\cite{Chivukula:2002ej}, it is expected that the unitarity
cancellations work approximately for a large number of sites in theory
space models.

We now will evaluate the sum rules~\eqref{eq:fermi-srs}
for processes involving the third family quarks and the $W$ and $Z$ bosons.
The KK-modes of the bottom and top quarks will be
denoted as $T^{(n)}$ and $B^{(n)}$. 
In a 5D model, there can be separate towers of vector-like KK-modes
for the left-handed and right-handed quarks but for notational simplicity, 
we will not introduce a different notation for these towers.
The $b$ quark is treated as massless everywhere and frequently 
axial couplings $g^A=\frac{1}{2}(g_R-g_L)$ are used.  

For the processes $W^+ W^- \to t\bar t$ and $W^+ W^- \to b\bar b$
the condition~\eqref{eq:fermi-lie} gives
\begin{subequations}\label{eq:tt-lie}
  \begin{align}
 -(g^{L/R}_{Wtb})^2+
     g_{ZWW}g_{ttZ}^{L/R}+ g_{WW\gamma}g_{tt\gamma}&=
    \sum_n\left[(g^{L/R}_{WtB^{(n)}})^2
      -g_{WWZ^{(n)}}g_{ttZ^{(n)}}^{L/R}\right]
\overset{\text{SM}}{=}0\\
 (g^{L/R}_{Wtb})^2+
    g_{ZWW}g_{bbZ}^{L/R}+ \ii g_{WW\gamma}g_{bb\gamma}&=-
    \sum_n\left[(g^{L/R}_{WT^{(n)}b})^2
      + g_{WWZ^{(n)}}g_{bbZ^{(n)}}^{L/R}\right]
\overset{\text{SM}}{=}0
\end{align}
\end{subequations}
The sign change in the relation for bottom quarks arises since a
different term in the Lie algebra~\eqref{eq:fermi-lie} contributes.  In
the following, we will concentrate on the sum rule for top quarks, the
case of bottom quarks is similar.  As indicated, the expressions on
the left hand side vanish if the tree level SM
values are inserted, reflecting the fact that these relations are a
consequence of gauge invariance alone, independent of the mechanism of
EWSB.  Note however, that
many top quark couplings are presently constrained only indirectly
by experiment, for instance by assuming unitarity of the 
Cabibbo-Kobayashi-Maskawa matrix.
In a Higgsless model the couplings of the
zero modes will in general receive corrections compared to the SM. 
In particular, as mentioned in section~\ref{sec:examples} one expects
modified couplings of the right-handed top quark, arising from localization
towards the EWSB brane in 5D models or from a large Yukawa
coupling on the $U(1)_Y$ site in theory space models.
 The condition
for the cancellation of the subleading
divergences~\eqref{eq:fermi-gold-KK} results in
\begin{equation}
\label{eq:tt-yukawa}
(g^{L}_{Wtb})^2
      + g_{ZWW}g_{ttZ}^A =
 \sum_{n}\Biggl[
 \tfrac{m_{B^{(n)}}}{m_t}g^R_{WtB^{(n)}}g^L_{WtB^{(n)}}
 -(g^{L}_{WtB^{(n)}})^2-
 g_{WWZ^{(n)}}g_{ttZ^{(n)}}^A    \Biggr]
\end{equation}
Subtracting the left- and right-handed versions of~\eqref{eq:tt-lie} 
 and  using~\eqref{eq:tt-yukawa} one can eliminate the $Ztt$ couplings :
\begin{equation}\label{eq:wtb-sr}
(g^{L}_{Wtb})^2+(g^{R}_{Wtb})^2= 
\sum_{n}\left[ 2\tfrac{m_{B^{(n)}}}{m_t}g^R_{WtB^{(n)}}g^L_{WtB^{(n)}}-(g^{L}_{WtB^{(n)}})^2-(g^{R}_{WtB^{(n)}})^2
 \right]
\end{equation}
Note that the disappearance of the couplings of the $Z$-KK-modes from
the sum rule implies that the unitarity cancellations \emph{cannot be
  achieved} by including only vector boson resonances and the presence
of fermion KK-modes is necessary.

Following~\cite{Birkedal:2004au} we should now proceed to obtain an upper
bound on the $g_{WtB^{(1)}}$ coupling by saturating the sum
rule~\eqref{eq:wtb-sr} by the first resonance.  However, in principle
individual terms of the sum~\eqref{eq:wtb-sr} can be negative if the left-and
right-handed couplings are very different.  On the other hand, for the higher
KK-modes this becomes increasingly unlikely because the positive contribution
is enhanced by the KK-mass.  Hence we expect that a bound similar to~\eqref{eq:bmp} can safely be obtained from the sum rule~\eqref{eq:wtb-sr} 
in a scenario with a non-degenerate KK-spectrum as in the theory space model
discussed in section~\ref{sec:examples}.  Introducing the notation $ g^{L}_{Wq
  Q^{(1)}}=\sqrt 2\sin\theta_Q g_{Wq Q^{(1)}}$ and $ g^{R}_{Wq Q^{(1)}}=\sqrt
2\cos\theta_Q g_{Wq Q^{(1)}}$ one then obtains
\begin{equation}\label{eq:wtB-bound}
  (g^{L}_{Wtb})^2\gtrsim  2g_{Wt B^{(1)}}^2
\left(\tfrac{ m_{B^{(1)}}}{m_t}\sin 2\theta_{B} -1\right)
\end{equation}
In contrast to the case of gauge boson scattering considered
in~\cite{Birkedal:2004au}, this bound is not model independent but
involves the relation of right-and left-handed couplings.  We will
solve the sum rules derived in this section under the assumption of a
non-degenerate spectrum and saturation by the first KK-level in the
next section.  For almost degenerate KK-excitations of the $b$ quark
as in some $SU(2)_L\otimes SU(2)_R\otimes U(1)$ 5D models, in
principle there can be cancellations among the degenerate modes and
more detailed knowledge of the chiral structure of the couplings would
be necessary to exploit the sum rules.

Turning now to the process $ZZ \to t\bar t$,
in this case the sum rule~\eqref{eq:fermi-lie} is satisfied trivially, 
since there is no coupling of three neutral gauge bosons, even involving
KK modes. The remaining relation~\eqref{eq:fermi-gold-KK} gives
\begin{equation}
\label{eq:ttzz-sr}
4 \left(g_{ttZ}^A\right)^2=-\sum_n 
\left[\left(g_{t T^{(n)}Z}^L\right)^2+
  \left(g_{t T^{(n)}Z}^R\right)^2-2\tfrac{m_{T^{(n)}}}{m_t}
g_{tT^{(n)}Z}^R g_{tT^{(n)}Z}^L
 \right] 
\overset{\text{SM}}{=}\frac{g^2}{4\cos^2 \theta_W}
\end{equation}
Finally, there are the sum rules for $W^+ Z\to \bar b t$ that
have a more involved form than those considered previously:
\begin{subequations}\label{eq:tb-sr}
\begin{gather}
\label{eq:tb-sr1}
\begin{split}
  (g_{ttZ}^{L/R}&-g_{bbZ}^{L/R}- g_{ZWW})g_{Wtb}^{L/R}\\
&=-\sum_{n}\Biggl[
 g_{t T^{(n)}Z}^{L/R}g_{WT^{(n)}b}^{L/R}
-g_{bB^{(n)}Z}^{L/R}g_{WtB^{(n)}}^{L/R}
- g_{ZWW^{(n)}}g_{W^{(n)}tb}^{L/R}\Biggr]
\overset{\text{SM}}{=}0
\end{split}\\
\begin{split}
&-\left(2 g_{ttZ}^A+\tfrac{m_Z^2}{2m_W^2} g _{ZWW} \right) g_{Wtb}^L  \\
&=\sum_{n}\Biggl[
( \tfrac{m_{T^{(n)}}}{m_t} g_{tT^{(n)}Z}^R - g_{tT^{(n)}Z}^L)g_{WT^{(n)}b}^L
+\tfrac{m_{B^{(n)}}}{m_t}g_{WtB^{(n)}}^R
g_{bB^{(n)}Z}^L
+\tfrac{1}{2}g_{W^{(n)}tb}^L g _{ZWW^{(n)}}
\Biggr] 
\overset{\text{SM}}{=}0
\end{split}
\end{gather}
\end{subequations}
where the second equation holds up to terms of the order $m_W^2/m^2_{W^{(n)}}$.
In this case, the condition with left- and right-handed couplings
exchanged gives an independent condition.
The only new couplings appearing in these relations are the
coupling of the $W$-KK modes $g_{W^{(n)}tb}$. 
Combining  both relations of~\eqref{eq:tb-sr}
results in the consistency relation
\begin{equation}\label{eq:consistency}
0\overset{\text{SM}}{=}  \sum_n \left(2\tfrac{m_{T^{(n)}}}{m_t}g_{tT^{(n)}Z}^R-g_{tT^{(n)}Z}^L\right)
g_{WT^{(n)}b}^L
+\left(2\tfrac{m_{B^{(n)}}}{m_t}g_{WtB^{(n)}}^R -g_{WtB^{(n)}}^L\right)
g_{bB^{(n)}Z}^L 
\end{equation}

The sum rules derived in this section are sufficient to ensure that
the matrix elements for four-particle processes with external SM
particles remain bounded at large energies.  For a fully consistent
model, in addition to the interaction considered
in~\eqref{eq:lagrangian} also coupling constants among the KK-modes have
to be taken into account, satisfying the appropriate sum rules to
cancel unitarity violations in the scattering amplitudes of the
KK-resonances~\cite{Csaki:2003dt}. Furthermore despite cancellation of
the terms growing with the energy, eventually partial wave unitarity
will be violated by the growing multiplicity of open
channels~\cite{SekharChivukula:2001hz,Papucci:2004ip}.

\section{Implications for collider phenomenology}
\label{sec:pheno}
Obtaining generic predictions from the unitarity sum rules derived in
the last section is less straightforward than in gauge boson scattering
considered in~\cite{Birkedal:2004au}. 
With some further input from model-building or some simplifying
assumptions, however, these relations can be useful to constrain
the interactions of the first KK-level  
 of a Higgsless model.  This is demonstrated in subsection~\ref{sec:couplings}
for a simple setup with a non-degenerate mass spectrum. 
In subsection~\ref{sec:cross-sections} we
 give examples for the high energy behavior of some four particle
 cross sections in this scenario and study the effects of varying
the KK-masses and coupling constants. 
In subsection~\ref{sec:collider} we
compare our scenario to top-sector signatures of Little Higgs models
and general vector resonances in models of strong EWSB.

\subsection{Minimal Higgsless scenario for the first KK-level}
\label{sec:couplings}
In the following all zero-mode couplings will be approximated by their
tree-level SM value (implicitly also done in~\cite{Birkedal:2004au}) and it is
assumed that the sum rules are saturated by the first resonance.
It is straightforward to allow for deviations of the zero mode couplings from
their SM values, we comment on this briefly below.  Almost degenerate
KK-resonances of the quarks will not be considered, as discussed in
section~\ref{sec:examples} this corresponds to theory space models of the type
considered in~\cite{Chivukula:2005bn} or the continuum limit as
in~\cite{Foadi:2004ps}.  Under these assumptions, the sum rules derived in
section~\ref{sec:sr} can be solved so that all four point amplitudes of the SM
fermions and gauge bosons remain bounded at high scattering energies.
A unitarization of four point amplitudes with external
particles from the first KK-level would require the inclusion of higher 
KK levels~\cite{Csaki:2003dt}.
 The scenario described in this section has been implemented
into the multi-purpose event-generator \texttt{O'Mega/WHIZARD}~\cite{OMega}.

In the following, the explicit analytic expressions of the
coupling constants in terms of masses and SM couplings
will be displayed only for the
case of equal left-and right handed couplings. As argued
in subsection~\ref{sec:cross-sections} below, the
deviations from this limit should remain small in order to keep
the cross sections significantly beyond the SM predictions in the
$m_H\to\infty$ limit up to $\sqrt s=5$ TeV. In our implementation in
\texttt{O'Mega}, the left-and right-handed couplings are kept as free
input parameters and the exact formulas resulting from the sum rules are
used. 
Ambiguities in the absolute signs of the coupling
constants will be fixed to satisfy the constraint~\eqref{eq:consistency}.

Considering~\eqref{eq:wtB-bound} in the vector-like limit $\theta_Q=\pi/4$
 and inserting the SM value $g^L_{Wtb}=g/\sqrt 2$ gives:
\begin{equation}\label{eq:wTb-hl}
g_{Wq Q^{(1)}}^{L/R}
\approx \frac{g}{2}\sqrt{\frac{ m_q}{m_{Q^{(1)}}}}
\end{equation}
Similarly, from the sum rules~\eqref{eq:tt-lie} from $W^+ W^-\to q\bar
q$ one obtains, truncating after the first KK-level and using the SM
value for the zero mode couplings:
\begin{equation}\label{eq:nc}
   g_{qq Z^{(1)}}^{L/R}=(-)^{\chi_q}
   \frac{(g_{WqQ^{(1)}}^{L/R})^2}{g_{WWZ^{(1)}}} 
=(-)^{\chi_q}\frac{\sqrt 3 g}{4}\frac{m_q m_{Z^{(1)}}}{m_W m_{Q^{(1)}}} 
\end{equation}
with $\chi_t=0$ and $\chi_b=1$.
The neutral current coupling of the quark resonances is given from 
\eqref{eq:ttzz-sr} as
\begin{equation}\label{eq:zTt-hl}
  g_{q Q^{(1)}Z}^{L/R}\approx
\sqrt{\frac{2m_q}{m_{Q^{(1)}}}} g^A_{qqZ}
= (-)^{\chi_q+1}\frac{g}{2\cos\theta_w}\sqrt{\frac{m_q}{2m_{Q^{(1)}}}}
\end{equation}
The remaining coupling $g_{W^{(1)}tb}$ can be fixed using~\eqref{eq:tb-sr1}:
\begin{equation}
\label{eq:gw1tb}
  g_{W^{(1)}tb}^{L/R}=\frac{1}{g_{ZWW^{(1)}}}( g_{t T^{(1)}Z}^{L/R}g_{WT^{(1)}b}^{L/R}-g_{bB^{(1)}Z}^{L/R}g_{WtB^{(1)}}^{L/R})
\approx
-\frac{g m_{W^{(1)}}}{2 m_{B^{(1)}}  m_W }\sqrt {\frac{3}{2} m_t m_b }
\end{equation}
where the last expression holds provided $m_{T^{(1)}}\approx m_{B^{(1)}}$. 
Finally, under the condition that the terms proportional to the KK-masses
dominate,  the consistency condition~\eqref{eq:consistency} simplifies to
\begin{equation}
  m_{T^{(1)}}g_{tT^{(1)}Z}^R g_{WT^{(1)}b}^L=-m_{B^{(1)}}g_{WtB^{(1)}}^R
g_{bB^{(1)}Z}^L 
\end{equation}
which is satisfied for our sign conventions used above. 
For deviations from the vector-like limit, this condition and
the one obtained by exchanging left- and right-handed couplings 
constrains the ratios of left-and right handed couplings, leaving us
with two additional free parameters.
Together with the triple
gauge boson couplings~\eqref{eq:bmp}, these results
provide a simple
description for the first KK-level of a Higgsless model.

Let us briefly discuss the impact of modified zero mode couplings.
Consider a non-vanishing right-handed $Wtb$ coupling, parameterized as
$  g^R_{Wtb}=\frac{g}{\sqrt 2}\epsilon_R$.
From~\eqref{eq:tt-lie},~\eqref{eq:wtb-sr} and~\eqref{eq:tb-sr1}
we obtain the relative changes in the 
couplings~\eqref{eq:wTb-hl},~\eqref{eq:nc} and~\eqref{eq:gw1tb} as
\begin{equation}
     \frac{\delta g_{Wt B^{(1)}}^{L/R}}{g_{Wt B^{(1)}}^{L/R}}
     \approx \frac{\epsilon_R^2}{2}\qquad , \qquad
   \frac{\delta g_{tt Z^{(1)}}^{R}}{g_{tt Z^{(1)}}^{R} }\approx
\frac{m_{B^{(1)}} }{m_t}\epsilon_R^2  \qquad , \qquad
     \frac{\delta g_{W^{(1)}tb}^{R}}{g_{W^{(1)}tb}^{R} } \approx
\frac{2 m_{B^{(1)}}}{\sqrt {m_t m_b}}\epsilon_R
\end{equation}
A non-vanishing right-handed $Wtb$ coupling therefore gives only small 
corrections to the couplings of the quark KK-modes while the
couplings of the $Z$ boson KK-modes can be enhanced moderatly
 for instance $\epsilon_R=5\%$ results in 
 $\delta g_{tt Z^{(1)}}^{R}/g_{tt Z^{(1)}}^{R}\approx 4\%$
for $m_{B^{(1)}}=3$ TeV.
In contrast, large changes can arise in the couplings of the KK-modes of the
$W$ boson.

We are now in the position to
determine the decay widths of the
KK particles.
The partial decay widths for $W^{(1)}\to  W Z$~\cite{Birkedal:2004au}
and $Z^{(1)}\to W^+ W^-$ are given by
\begin{equation}
\begin{aligned}
  \Gamma_{W^{(1)}\to WZ}&\approx\frac{g_{W^{(1)}WZ}^2}{48\pi}
\frac{m^5_{W^{(1)}}}{4 m_W^2m_Z^2}
\lesssim \frac{\alpha_{\qed} m_{W^{(1)}}^3}{144 \sin^2\theta_w m_W^2}\\
 \Gamma_{Z^{(1)}\to WW}&\approx\frac{g_{WWZ^{(1)}}^2}{48\pi}
\frac{m^5_{Z^{(1)}}}{4 m_W^4}
\lesssim \frac{\alpha_{\qed} m_{Z^{(1)}}^3}{144 \sin^2\theta_w m_W^2}
\end{aligned}
\end{equation}
For $m_{W^{(1)}}=m_{Z^{(1)}}=700$ GeV the numerical value is approximately $13$ GeV.
For vector-like couplings, the partial decay width of the heavy $Z$
into top-quarks is given by
\begin{equation}
\label{eq:br_z1tt}
  \Gamma_{Z^{(1)}\to t\bar t}
\approx\frac{g^2_{tt Z^{(1)}} m_{Z^{(1)}}}{4\pi}
\approx\frac{3\, \alpha_{\qed} m^3_{Z^{(1)}}}{16 \sin^2\theta_w m_W^2}
 \left(\frac{m_t}{m_{B^{(1)}}}\right)^2
=27\left(\frac{m_t}{m_{B^{(1)}}}\right)^2 \Gamma_{Z^{(1)}\to W^+ W^-}
\end{equation}
For $m_{B^{(1)}}=2.5$ TeV 
 the numerical value is about $14\%$ of the $Z^{(1)}\to W^+ W^-$ branching ratio but for $m_{B^{(1)}}=1$ TeV it becomes as large as $86\%$. 
As can be seen from~\eqref{eq:gw1tb},
the fermionic branching ratio of the $W^{(1)}$ $\Gamma_{W^{(1)}\to t\bar b}$
is suppressed compared to that of the $Z^{(1)}$ by a factor $m_b/m_t$.

The partial decay widths of the heavy quarks into SM particles are
\begin{equation}
\label{eq:width_q1}
  \begin{aligned}
 \Gamma_{Q^{(1)}\to q Z}&\approx\frac{g_{Q^{(1)}qZ}^2}{16\pi} \frac{m_{Q^{(1)}}^3}{m_Z^2}=
\frac{\alpha_{\qed}}{32\sin^2\theta_w\cos^2\theta_w}\frac{m_qm_{Q^{(1)}}^2}{m_Z^2}
\\
 \Gamma_{Q^{(1)}\to q W}&\approx\frac{g_{Q^{(1)}qW}^2}{16\pi} \frac{m_{Q^{(1)}}^3}{m_W^2}
=\frac{\alpha_{\qed}}{16\sin^2\theta_w}\frac{m_qm_{Q^{(1)}}^2}{m_W^2}
  \end{aligned}
\end{equation}
The dominant decay channels are hence the neutral current 
decay for the $T^{(1)}$ and the charged current decay  for the 
$B^{(1)}$ that are not suppressed by the
small $b$-quark mass.  For $m_{T^{(1)}}=m_{B^{(1)}}= 1$ TeV we find
$\Gamma_{T^{(1)}\to t Z}\approx 30$ GeV and $\Gamma_{B^{(1)}\to t
  W}\approx 60$ GeV.  The dependence on $m_Q^2$ leads to a rapid
growth with the mass, for instance $m_{T^{(1)}}=3$ TeV and
$m_{B^{(1)}}=2.5$ TeV lead to $\Gamma_{T^{(1)}\to t Z}\approx 270$
GeV and $\Gamma_{B^{(1)}\to t W}\approx 380$ GeV.  Note that for such
large masses also decays like $Q^{(1)}\to Z^{(1)} q$ are kinematically
allowed.  In a 5D theory with a flat extra dimension compactified on an
orbifold, the corresponding KK-number conserving coupling constants
are of the same order of magnitude as the zero mode couplings, whereas 
couplings involving a single KK mode
are suppressed since they are generated only by KK-parity violating
boundary terms.  
While we have not worked out the unitarity
constraints for the KK-number conserving coupling constants, we
expect they are not suppressed by factors
$\mathcal{O}(\sqrt{(m_q/m_Q)})$.  We estimate the order of magnitude for
the partial decay width for these channels by $\Gamma_{T^{(1)}\to t
  Z^{(1)}}\sim (m_{T^{(1)}} m_Z^2)/(m_t m_{Z^{(1)}}^2)
\Gamma_{T^{(1)}\to t Z}\sim 0.3\Gamma_{T^{(1)}\to t Z}$ for $m_{T^{(1)}}= 3$ TeV.
 While
this simple argument suggests that the
decay to one KK-mode and a zero mode is subdominant, this point can 
only be settled within a concrete model.

\subsection{High energy-behavior of cross sections }
\label{sec:cross-sections}
We now give results for the cross sections $W^+ W^-\to t\bar t$ and $Z Z\to t\bar t$ in the scenario discussed in the previous subsection. 
This serves on one hand to demonstrate that our choice of coupling
constants indeed implements the required unitarity cancellations
and leads to decreasing cross sections at high energies. 
On the other hand, we address the question whether
one can safely raise the KK-masses of the third family quarks near the AC-bound
while significantly improving the high energy behavior
compared to the SM in the limit of an infinite Higgs mass.
We also discuss whether the couplings can be raised significantly compared
to the values discussed above, either by deviating from  vector-like couplings
or by violating the sum rules.

\begin{figure}[htbp]
 \begin{center}
   \includegraphics[width=\graphicswidth]{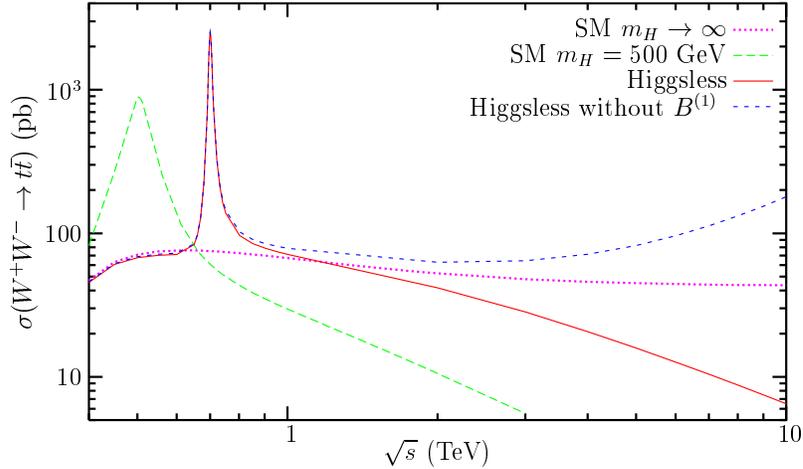}
    \caption{Cross section for $W^+ W^-\to t\bar t$ in the
      SM with a Higgs resonance, the SM in the $m_H\to\infty$
      limit and the Higgsless scenario with $m_Z^{(1)}=700$ GeV and
      $m_{B^{(1)}}=2.5$ TeV}
\label{fig:wwtt}
\end{center}
\end{figure}
In figure~\ref{fig:wwtt} the cross section for $W^+ W^-\to
t\bar t$ is shown in various scenarios. As expected, in the SM in the limit of
an infinite Higgs mass the cross section tends to a constant at high
energies, corresponding to a scattering matrix element growing
linearly with the energy. Both in the SM with a Higgs resonance (for
comparison with the Higgsless model, a rather heavy
Higgs with $m_H=500$ GeV is shown  in this plot, but the high energy limit is
the same for a lighter Higgs) and in the Higgsless scenario the
scattering matrix element is bounded at large energies and the cross
section decreases.  However, in the Higgsless scenario this decrease 
sets in at much higher energies than in the SM.  It
has been checked, that this behavior is not improved for smaller
$m_{Z^{(1)}}$.  As emphasized previously, the improved high energy
behavior in the Higgsless scenario is due to the heavy $B$ quark.
Indeed, as can be seen in figure~\ref{fig:wwtt} and in agreement with
the sum rule~\eqref{eq:tt-lie}, the inclusion of a single $Z$
resonance without a heavy $B$ rather \emph{destroys} the unitarity
cancellations present in the SM and leads to a growing cross section
at high energies (for comparison the couplings of the $Z^{(1)}$ have
been taken as in~\eqref{eq:nc} with $m_{B^{(1)}}=2.5$ TeV).

\begin{figure}[htbp]
 \begin{center}
   \includegraphics[width=\graphicswidth]{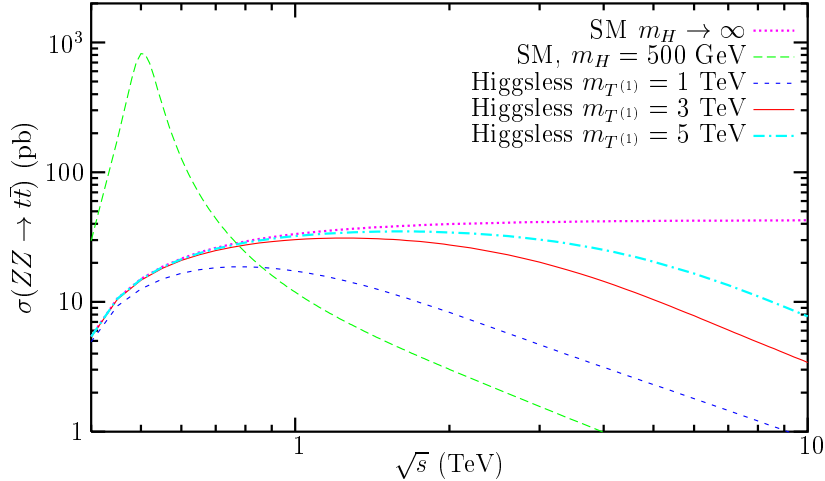}
    \caption{Cross section for $ZZ\to t\bar t$ in the
      SM with a Higgs resonance, the SM in the $m_H\to\infty$
      limit and the Higgsless SM for three different masses of the heavy top.}
\label{fig:zztt}
\end{center}
\end{figure}
Figure~\ref{fig:zztt} shows the cross section for the process 
$ZZ\to t\bar t$ for three different values of $m_{T^{(1)}}$.
 Here no resonance appears in the Higgsless
model so the unitarization is entirely due to the top quark KK-mode.
It can be seen that that the numerical value of the cross section at
large energies is determined by the mass of the $T^{(1)}$ since the
unitarity cancellations become effective earlier on for a lower mass.
For $m_{T^{(1)}}=1$ TeV the cross section is suppressed considerable
compared to the $m_H\to \infty$ limit of the SM already for $\sqrt s$=
2 TeV. At $\sqrt s$= 3.5 TeV the cross section remains about $20\%$
below the infinite Higgs mass limit for masses up to $5$ TeV.  The
situation is similar for $W^+ W^-\to t\bar t$.  While a partial wave
analysis as
in~\cite{SekharChivukula:2001hz,Chivukula:2002ej,Papucci:2004ip} is
required to arrive at definite conclusions, this result indicates that
it should be possible to raise the mass of the KK-modes of the top and
bottom quarks as seems to be required by low energy constraints
without running in conflict with unitarity.

\begin{figure}[htbp]
 \begin{center}
     \includegraphics[width=\graphicswidth]{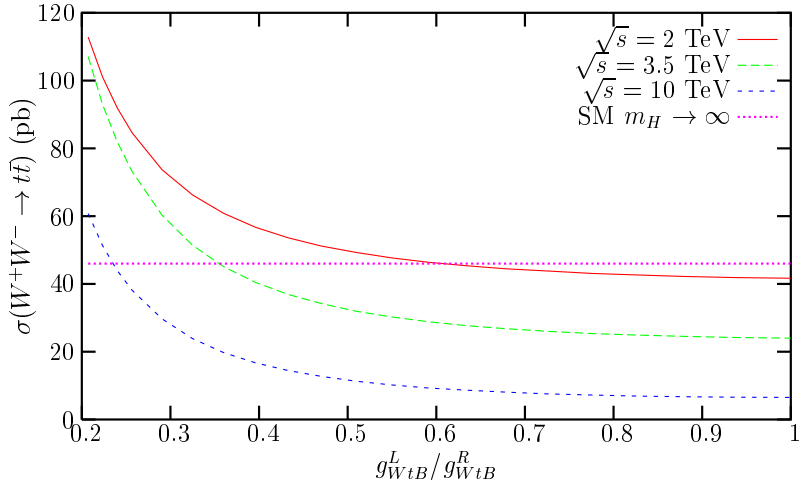}
    \caption{Cross section for $W W\to t\bar t$ for a varying
       ratio of left-and right handed $WtB^{(1)}$ couplings while keeping
      the bound~\eqref{eq:wtB-bound} saturated. }
\label{fig:th_lr}
\end{center}
\end{figure}

The explicit expressions for the coupling constants
like~\eqref{eq:wTb-hl} and~\eqref{eq:zTt-hl} hold only if the
difference between left- and right-handed couplings is not too large.
One observes from~\eqref{eq:wtB-bound} that
in principle the coupling constants can be enhanced by a factor $(\sin
2\theta_B -m_t/m_{B^{(1)}})^{-1}$.  On the other hand, this
also increases the cross section so that  the demand to do better than the SM
in the infinite Higgs mass limit at $\sqrt s= \Lambda_t= 3.5$ TeV
places a bound on the ratio
$g_{WtB^{(1)}}^L/g_{WtB^{(1)}}^R=\tan \theta_B$. From
figure~\ref{fig:th_lr}
one can see that this implies
$\tan\theta_B \gtrsim 0.4$, corresponding to an enhancement
of $g_{WtB^{(1)}}$ by a factor of $1.6$ compared to~\eqref{eq:wTb-hl}.
Therefore the relations~\eqref{eq:wTb-hl} and~\eqref{eq:zTt-hl} will
not be corrected by large numbers, if one insists on a better high
energy behavior than in the SM with $m_H\to\infty$.

\begin{figure}[t]
 \begin{center}
  \includegraphics[width=\graphicswidth]{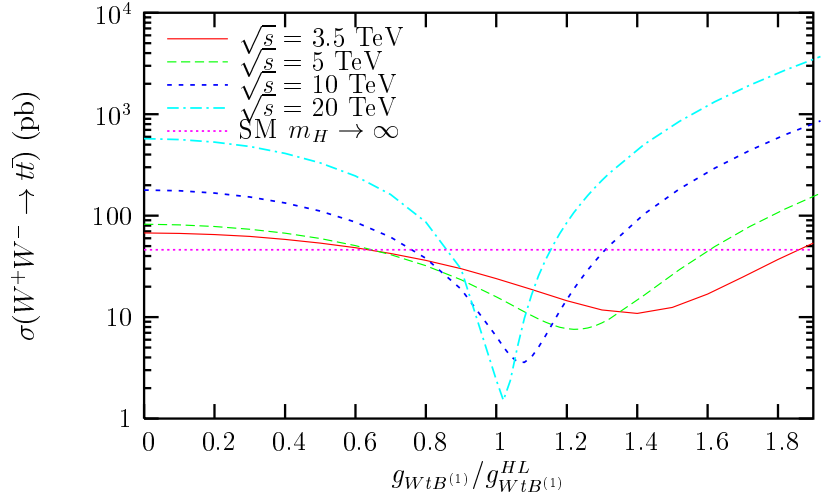}
    \caption{Dependence of the $W W\to t\bar t$ cross section
      on the  $WtB^{(1)}$ coupling
      where $g^{HL}_{WtB^{(1)}}$ denotes the value~\eqref{eq:wTb-hl}. 
      The remaining couplings
      are kept fixed.}
\label{fig:gwtb1}
\end{center}
\end{figure}

Another interesting question concerns the effect of  small violations of the
unitarity sum rules on the high energy behavior of the cross
section.  For the example of the $W^+W^-\to t\bar t$ cross section, 
figure~\ref{fig:gwtb1} shows the effect of varying the
$WtB^{(1)}$ coupling while keeping the remaining couplings fixed.
As a first observation, one notes that  the coupling constant
derived in subsection~\ref{sec:couplings} 
indeed is  `optimal' in the sense that the cross section
has the smallest value in the high energy limit. For smaller energies, however, the `optimal' coupling is shifted to larger values, indicating
 that the cancellations
induced by the heavy $B$ quark become more effective earlier on for
an increased coupling, while in the high energy limits the cancellations
are spoiled.
Depending on the cutoff scale $\Lambda\sim 5$-$10$ TeV
 of the theory induced by perturbativity in gauge boson scattering, 
the coupling $g_{WtB}$ might be raised at most to $1.2$-$1.4$ times the value
 inferred from~\eqref{eq:wTb-hl}.

\subsection{Comparison with other EWSB scenarios}
\label{sec:collider}
Let us now compare our scenario with the collider phenomenology of
fermion or vector boson resonances in the top sector that have been
considered in the context of Little Higgs
models~\cite{Han:2003wu,Azuelos:2004dm} or 
 strong EWSB~\cite{RuizMorales:1999kz,Han:2003pu}. We
first turn to the top-quark signals of the gauge boson KK-modes
before we discuss the KK-modes of the third family quarks.

Signals of a $1$ TeV vector resonance 
in the vector boson fusion process $W^+ W^- \bar t t$ were studied
in~\cite{RuizMorales:1999kz}
for an $e^+ e^-$ linear collider
operating at $\sqrt s=1.5$ TeV.   From the results of the second
reference in~\cite{RuizMorales:1999kz} one finds that
this process 
is suitable for probing the partial widths in the
region $\Gamma_{Z^{(1)}\to W^+ W^-} \gtrsim 10$ TeV and
$\Gamma_{Z^{(1)}\to t\bar t} \gtrsim 2.4$ TeV with a precision of
$13\%$ for a luminosity of $200\, \text{fb}^{-1}$.  The bosonic decay
width is of the expected magnitude in our scenario, for the fermionic
decay width of the $Z^{(1)}$~\eqref{eq:br_z1tt}, this region
corresponds to masses $m_{B^{(1)}}\lesssim 2.2$ TeV.  A high energy
linear collider therefore should be suitable to probe the coupling of
the third family quarks to the KK-modes of the gauge bosons and might
even provide a check on the unitarity sum rules.

Concerning hadron collider signatures, it was found that at the LHC the
detection of resonances in $W^+ W^- \to \bar t t$ is overwhelmed by QCD
background~\cite{Han:2003pu}.  Recently, the second reference
in~\cite{Han:2003pu} considered a different scenario with a neutral gauge
boson coupling predominantly to the third generation of quarks with the
coupling to gauge bosons and light fermions suppressed.  The associated
production of a heavy vector boson with $b$ or $t$ quarks, for instance in
single top production $W b\to t Z^{(1)}$ or in bottom or top pair
production $g g \to t\bar t Z^{(1)}$ has been found to be
the most useful signature.  For a partial decay width $\Gamma_{Z^{(1)}\to t
  \bar t}= 63$ GeV this reference finds a $5\sigma$ significance at the LHC
for masses up to $m_{Z^{(1)}}=2$ TeV.  In our scenario the partial decay width
of the heavy $Z$ to top quarks is only about $4\%-17\%$ of the value
considered in~\cite{Han:2003pu}, depending on $m_{B^{(1)}}$.  However, for
$m_{Z^{(1)}}< 1$ TeV---as expected from unitarity in gauge boson
scattering---the cross sections of the channels considered
in~\cite{Han:2003pu} grow rapidly so one expects that they are useful also in
the context of Higgsless models.  Clearly, a more careful study in the
scenario described in~\ref{sec:couplings} is needed to arrive at definite
conclusions, for instance the total cross section for $W^+ W^-\to t\bar t$ at
high energies (c.f. figure~\ref{fig:wwtt}) is by construction much smaller
than in the scenarios with a single vector resonance considered
in~\cite{RuizMorales:1999kz,Han:2003pu}.

While the detection of the KK-resonances of the gauge bosons would provide
evidence for the nature of the mechanism of gauge symmetry breaking, 
the KK-resonances of the top and bottom quark are  connected to the
mechanism of mass generation of the quarks
and might help to distinguish 5D from theory space models.
The production of heavy top quarks at the LHC has
been studied recently in the context of Little Higgs models~\cite{Han:2003wu,Azuelos:2004dm}. 
In the so called Littlest
Higgs model~\cite{Arkani-Hamed:2002qy},  
the couplings of the heavy top $T$  are suppressed compared to that
of the top quark by a mixing angle given approximately by~\cite{Han:2003wu} 
 $\lambda_1 m_t/(\lambda_2m_T)$ where the ratio of Yukawa couplings $\lambda_1$ and $\lambda_2$ is usually
taken to be of the order one.
For $\lambda_1/\lambda_2=1$, the discovery reach of the LHC for the heavy top
has been estimated~\cite{Azuelos:2004dm} as $m_T=2$ TeV for the  production
in $W$-$b$ fusion $q b\to q' T$ and the decay channel $T\to Wb$.
In comparison to our result~\eqref{eq:width_q1}, the total decay width of the heavy top in the Littlest Higgs model
is given by~\cite{Han:2003wu}
\begin{equation}
  \Gamma_T^{\text{LH}}\approx \frac{\alpha_\qed m_T m_t^2}{8\sin^2\theta_w
  m_W^2}\left(\frac{\lambda_1}{\lambda_2}\right)^2
  = \frac{4 m
_t}{m_T} \left(\frac{\lambda_1}{\lambda_2}\right)^2 
  \Gamma^{\text{HL}}_{T^{(1)}\to t Z}
\end{equation}
with branching ratios given by~\cite{Han:2003wu} $\text{Br}(T\to b W)=50\%$ and
$\text{Br}(T\to t Z)=\text{Br}(T\to t H)=25\%$. Recall
from~\eqref{eq:width_q1} that the decay $T^{(1)}\to b W$ is suppressed by a
factor $2 m_b/m_t\sim 5\%$ in the Higgsless scenario. Thus for 
 $m_{T^{(1)}}=1$ TeV the heavy top  has $\Gamma_T^{LH}=22$ GeV in the
Littlest Higgs model and is moderately narrower than in the Higgsless scenario.
In contrast, for $m_T=3$ TeV the heavy top remains relatively narrow in the
little Higgs model with $\Gamma_T^{LH}=62$ TeV while the width is over four
times larger in the Higgsless scenario.
On the other hand, the charged coupling constants involved in the
production of the heavy top are given by $g_{WTb}^{L}=\frac{g}{\sqrt 2}
\frac{m_t\lambda_1}{m_T\lambda_2}$ compared to our result~\eqref{eq:wTb-hl}.
Noting that the relevant quantity is the sum of the squared left-and
right-handed couplings, the production cross section in the Littlest
Higgs model and the Higgsless scenario
will be related  by 
\begin{equation}
  \sigma_{\text{LH}}(Wb\to T)=\frac{m_t^2 }{ m_b m_T} \left(\frac{\lambda_1}{\lambda_2}\right)^2  \sigma_{\text{HL}}(Wb\to T^{(1)})
\end{equation}
For $m_T=1$ TeV, the cross section in the Higgsless scenario corresponds
to that in the Little Higgs scenario for $\lambda_2=2.7\lambda_1$. 
In this mass region, the phenomenology of the heavy top hence is 
similar to that in the Littlest Higgs model, albeit for a slightly
pessimistic point in parameter space. For  
$m_T=3$ TeV this improves to  $\lambda_2=1.6\lambda_1$ but in this mass region
the heavy top is already a rather broad resonance so its phenomenology
will be even more challenging than in the Littlest Higgs model.

The heavy bottom quark is a feature distinguishing the Higgsless scenario from
(minimal implementations of) Little Higgs models.  Here the charged current
coupling is enhanced by the top quark mass, unfortunately the production in
the $s$-channel process $q\bar q\to W^-\to B^{(1)} \bar t$ is kinematically
suppressed at LHC energies.  Another possible production channel, neutral
current $b Z$ fusion $q b \to q B^{(1)}$ suffers from the suppressed neutral
current coupling, so the situation is similar as for the heavy top quark.

One could also consider the production of the heavy quarks by QCD
processes. 
In Little Higgs models, strong pair production of the heavy top quark
 has found to be kinematically suppressed compared to the weak process
of single $T$ production~\cite{Han:2003wu}.
Strong effects might be more relevant in higher dimensional models
where there are also effects of the KK-modes of the gluons (see
e.g.~\cite{Agashe:2003zs}) that, however, are not special to Higgsless
models and cannot be constrained in our approach.
If the KK-modes of the top and bottom quark evade direct detection
at the LHC, effects of
quark mixing with the KK-modes may only be observable by indirect effects
on the top quark gauge couplings~\cite{delAguila:2000kb}. 

\section{Summary and outlook}
In the spirit of a recent analysis of gauge boson scattering in
generic Higgsless models~\cite{Birkedal:2004au}, we have constrained
the interactions in the top sector of Higgsless models by unitarity
sum rules.  While the KK-resonances of the bottom and top quarks are
essential for good high energy behavior of scattering amplitudes
involving the top quark, electroweak precision constraints in a 5D
Higgsless model suggest that they must be significantly heavier than
the gauge boson KK-modes~\cite{Cacciapaglia:2004rb}.  In
section~\ref{sec:examples} we have seen that this can be achieved more
naturally in theory-space models than in 5D models.

Although a larger number of coupling constants is involved 
compared to gauge boson scattering discussed
in~\cite{Birkedal:2004au}, the
sum rules presented in section~\ref{sec:sr} 
 constrain the parameter
space significantly.  In section~\ref{sec:pheno} we have solved them
in the approximation that only the first KK-level contributes and for
a non-degenerate mass spectrum as in the theory space model discussed
in section~\ref{sec:examples}.  A numerical analysis of the cross
section for $W^+ W^-\to t\bar t $ in this scenario has shown that the
coupling constants can only deviate from the values given
in section~\ref{sec:couplings} by  a  limited amount 
if the high energy behavior is to
improve significantly compared to the SM in the $m_H\to\infty$ limit.
It will be interesting to compare the top sector of a realistic theory
space or higher dimensional Higgsless model to the scenario discussed
in section~\ref{sec:pheno}. 

The collider phenomenology of the gauge boson KK-modes in our setup
has some features in common with certain models of strong EWSB where
vector resonances couple predominantly to the third
generation~\cite{RuizMorales:1999kz,Han:2003pu} while the feature of a
excited top quark with mass $m_T=1$-$3$ TeV is shared by Little Higgs
models.  Comparison with results
of~\cite{RuizMorales:1999kz,Han:2003pu} suggests that the coupling of
the first $Z$ boson KK-mode to the top quark can be probed in
$t\bar t$ production via vector boson fusion at a high
energy linear collider~\cite{RuizMorales:1999kz} and in associated
production with top quarks at the LHC~\cite{Han:2003pu}.  The
phenomenology of the top and bottom quark KK-modes is more challenging
since the charged current interactions of the heavy top and the
neutral current interactions of the heavy bottom are suppressed by the
small mass of the bottom quark.  If the top and bottom quark KK-modes
indeed have to be significantly heavier than that of the gauge bosons,
as suggested in~\cite{Cacciapaglia:2004rb}, they will be difficult to
detect at the LHC.  The scenario described in section~\ref{sec:pheno}
has been implemented into the multi-purpose event-generator
\texttt{O'Mega/WHIZARD}, allowing for more detailed phenomenological
studies in the future.

\paragraph*{Note added:}
After the submission of this paper, ref~\cite{Cacciapaglia:2005pa}
appeared that discusses a model for the third family quarks based on
two slices of $AdS_5$ space. 
In this setup scalar "top-pions" appear and
the KK-resonances of the top and bottom quark are in the $6-7$
TeV region. The 
top-sector is either strongly coupled or a  a "Top-Higgs" is introduced,
restoring unitarity in $W^+W^-\to t\bar t$ scattering.
In contrast, the present work considered a top sector that remains 
perturbative without introduction of a scalar boson.

\section*{Acknowledgments}
This work has been supported
by the Deutsche Forschungsgemeinschaft through the
Gra\-du\-ier\-ten\-kolleg `Eichtheorien' at Mainz University.


\end{document}